\documentclass[aps,prl,floatfix,superscriptaddress,reprint]{revtex4-1} 
\usepackage{graphicx}
\usepackage{amsmath,amsfonts,amssymb}
\usepackage{color}
\usepackage{pdfpages}

\makeatletter
\AtBeginDocument{\let\LS@rot\@undefined}
\makeatother

\begin{document}

\title{Probing $\alpha$-RuCl$_3$ Beyond Magnetic Order: Effects of Temperature and Magnetic Field}
\date{\today}
\begin{abstract}
Recent studies have brought $\alpha$-RuCl$_3$ to the
forefront of experimental searches for materials realizing Kitaev spin-liquid
physics. This material exhibits strongly anisotropic exchange interactions afforded by
the spin-orbit coupling of the 4d Ru centers. We investigate the
dynamical response at finite temperature and magnetic
field for a realistic model of the magnetic interactions in $\alpha$-RuCl$_3$.
 These regimes are thought to host unconventional paramagnetic
 states that emerge from the suppression of magnetic order.
Using exact diagonalization calculations
of the quantum model complemented by semi-classical analysis,
we find a very rich evolution of the  spin dynamics as
the applied field suppresses the zigzag order and stabilizes a
quantum paramagnetic state that is adiabatically connected to the fully polarized state at high fields.
 At finite temperature, we observe large redistributions of
spectral weight that can be attributed to the anisotropic frustration of the
model. These results are compared to recent experiments, and
provide a roadmap for further studies of
these regimes.\end{abstract}

\author{Stephen M. Winter*}
\affiliation{Institut f\"ur Theoretische Physik, Goethe-Universit\"at Frankfurt,
Max-von-Laue-Strasse 1, 60438 Frankfurt am Main, Germany}
\author{Kira Riedl}
\affiliation{Institut f\"ur Theoretische Physik, Goethe-Universit\"at Frankfurt,
Max-von-Laue-Strasse 1, 60438 Frankfurt am Main, Germany}
\author{David Kaib}
\affiliation{Institut f\"ur Theoretische Physik, Goethe-Universit\"at Frankfurt,
Max-von-Laue-Strasse 1, 60438 Frankfurt am Main, Germany}
\author{Radu Coldea}
\affiliation{Clarendon Laboratory, University of Oxford, Parks Road, Oxford OX1 3PU, United Kingdom}
\author{Roser Valent{\'\i}}
\affiliation{Institut f\"ur Theoretische Physik, Goethe-Universit\"at Frankfurt,
Max-von-Laue-Strasse 1, 60438 Frankfurt am Main, Germany}

\maketitle

\begin{figure}
\includegraphics[width=\linewidth]{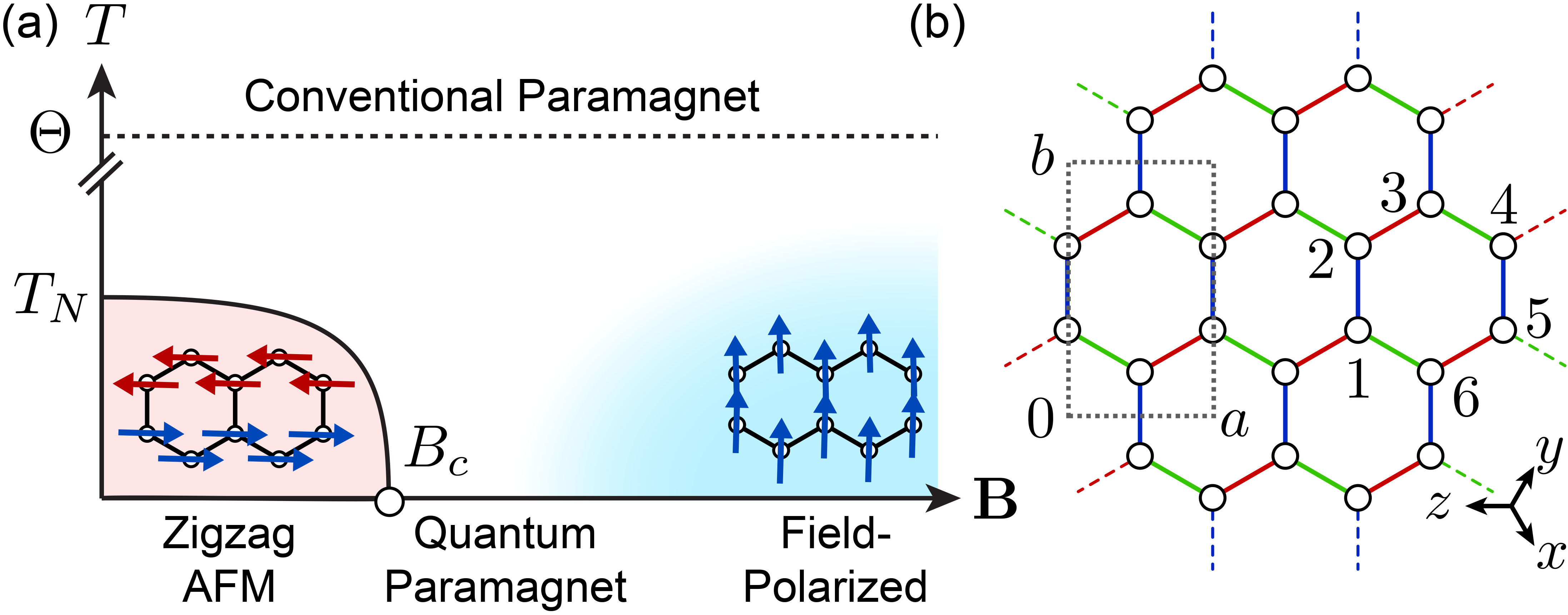}
\caption{(a) Schematic phase diagram of the model Hamiltonian (1) for $\alpha$-RuCl$_3$ at finite $T$ and $B$. $T_N$ is the N\'eel temperature, and $\Theta$ is the Curie-Weiss constant. The variable blue color shading indicates a crossover to the high-field regime. (b) 24-site cluster employed in ED calculations showing the orientation of the cubic $x,y,z$ axes, and $C2/m$ unit cell. The crystallographic axes correspond to $a = [11\bar{2}]$, $b=[1\bar{1}0]$, and $c^*=[111]$ in cubic coordinates. The numbers label sites defining the $\mathbb{Z}_2$ flux operator $\hat W_p$. Nearest neighbour X-,Y-, and Z-bonds are red, green, and blue, respectively. }
\label{fig-lat}
\end{figure}

{\it Introduction $-$} The honeycomb magnet $\alpha$-RuCl$_3$ has
recently received significant attention, in view of the ongoing search for
exotic magnetic states in real systems
\cite{0034-4885-80-1-016502,Norman2016QSL,RevModPhys.89.025003,trebst2017kitaev,schaffer2016recent,rau2016spin,hermanns2017physics,winter2017models}. This material has
anisotropic and frustrated magnetic interactions, which have been discussed in
the context of Kitaev's celebrated honeycomb model \cite{kitaev2006anyons}. The
ground state of this model is a gapless $\mathbb{Z}_2$ spin liquid that is
stabilized by bond-dependent coupling described by $\mathcal{H} = K_1\sum_{\langle
ij \rangle} S_i^\gamma S_j^\gamma$. Here $\gamma = \{x,y,z\}$ for the three
bonds emerging from each lattice site (Fig.~\ref{fig-lat}b). It has been
proposed that such interactions with $K_1<0$ can arise
\cite{Jackeli2009,chaloupka2013zigzag,PhysRevLett.112.077204,rau2014trigonal} from a delicate
balance of spin-orbit coupling (SOC), Hund's coupling, and crystal-field
splitting (CFS) that may be approximated in $\alpha$-RuCl$_3$ \cite{PhysRevB.90.041112,PhysRevB.91.241110}. As a result,
recent experiments
\cite{banerjee2016proximate,banerjee2016neutron,do2017incarnation,nasu2016fermionic}
have been discussed in the language of static fluxes and Majorana
spinons, which represent the exact excitations of the Kitaev spin liquid (KSL)
\cite{kitaev2006anyons,PhysRevB.92.115127,PhysRevLett.112.207203}. In practice,
however, the zero field ground state of $\alpha$-RuCl$_3$ exhibits zigzag
antiferromagnetic order \cite{PhysRevB.91.144420,PhysRevB.92.235119}
(Fig.~\ref{fig-lat}a), suggesting deviations from the interactions of the pure
Kitaev model. 
The specific nature of these deviations has been heavily discussed
\cite{PhysRevB.93.155143,PhysRevB.91.241110,
yadav2016kitaev,PhysRevB.93.214431,hou2016unveiling}, with most recent works
agreeing additional large
anisotropic couplings and long-range exchange likely stabilize magnetic order
\cite{winter2017breakdown,ybknew,yadav2016kitaev,PhysRevB.93.214431,catuneanu2017realizing}. Understanding the role of these interactions in the dynamic response remains a key challenge.

Dynamical probes, such as inelastic neutron scattering
\cite{banerjee2016proximate,banerjee2016neutron,ranneutron,do2017incarnation} (INS) and electron
spin resonance
\cite{little2017antiferromagnetic,wang2017magnetic,ponomaryov2017direct} (ESR),
have observed an unconventional continuum of magnetic excitations that coexist
with magnons below $T_N \sim$ 7 K. The identity of the continuum has captured significant focus, as
the connection to the Kitaev model remains an open question. Such continua may arise generically in the presence of bond-dependent anisotropic couplings \cite{winter2017breakdown}.
Recent interest
has therefore turned toward regimes where the suppression of zigzag order may
reveal the underlying character of the continuum (Fig.~\ref{fig-lat}a). For
example, order is suppressed by a small in-plane field of 
$B_c \sim 7$ T, giving rise to a much-discussed quantum paramagnetic state
\cite{baek2017observation,zheng2017gapless,hentrich2017large,wolter2017field,PhysRevB.95.180411,PhysRevLett.118.187203,Kimchi01}. Such behaviour may be analogous to the response of the 3D iridates $\beta,\gamma$-Li$_2$IrO$_3$ \cite{ruizarxiv,modicncomms}.
Finally, significant spin correlations persist in $\alpha$-RuCl$_3$ well above $T_N \sim$ 7 K,
suggesting a possible unconventional paramagnetic phase at intermediate
temperatures \cite{do2017incarnation,samarakoon2017comprehensive}. 
 In this
work, we discuss the physics in these regimes for a realistic model Hamiltonian for $\alpha$-RuCl$_3$ proposed in
\cite{winter2017breakdown} and compare with the available experimental observations.

{\it Model $-$} We focus on a simplified $C_3$-symmetric four-parameter model that has been shown to reproduce many aspects of the inelastic neutron scattering in the ordered phase at low temperature and zero field  \cite{winter2017breakdown}. Specifically: 
\begin{align}\nonumber
\mathcal{H} =& \  \sum_{\langle ij \rangle} J_1 \ \mathbf{S}_i \cdot \mathbf{S}_j + K_1 S_i^\gamma S_j^\gamma + \Gamma_1 \left( S_i^\alpha S_j^\beta + S_i^\beta S_j^\alpha\right) \\
& \ + \sum_{\langle \langle \langle ij \rangle \rangle \rangle} J_3 \ \mathbf{S}_i \cdot \mathbf{S}_j -\mu_B \sum_{i} \mathbf{B}\cdot \mathbf{g} \cdot \mathbf{S}_i
\end{align}
with nearest neighbour interactions $J_1 = -0.5$, $K_1 = -5.0$, and $\Gamma_1 = +2.5$ meV and third neighbour interaction $J_3 = +0.5$ meV. The pure Kitaev model corresponds to $J_1 = \Gamma_1 = J_3 = 0$. Here, $\mathbf{g}$ is the anisotropic $g$-tensor. 
%Based on CFS parameters computed from DFT \cite{PhysRevB.93.214431}, we have estimated $g_{c^*} \approx 1.3$ and $g_{ab} \approx 2.3$.
In the calculations we used $g_{c^*} = 1.3$ and $g_{ab} = 2.3$; these values are consistent with the range of previous theoretical estimates for $\alpha$-RuCl$_3$ \cite{PhysRevB.94.064435,yadav2016kitaev}, and experimental values for similar compounds \cite{stanko1973trigonal,jarrett1957paramagnetic,pedersen2016iridates}. 
We note that this simplified model underestimates the zero-field gap for excitations \cite{banerjee2016proximate,ranneutron,little2017antiferromagnetic}, which may be related to a weak breaking of $C_3$ symmetry in actual samples \cite{PhysRevB.92.235119}, or small additional interactions \cite{yadav2016kitaev,PhysRevB.93.214431}. 

\begin{figure}
\includegraphics[width=\linewidth]{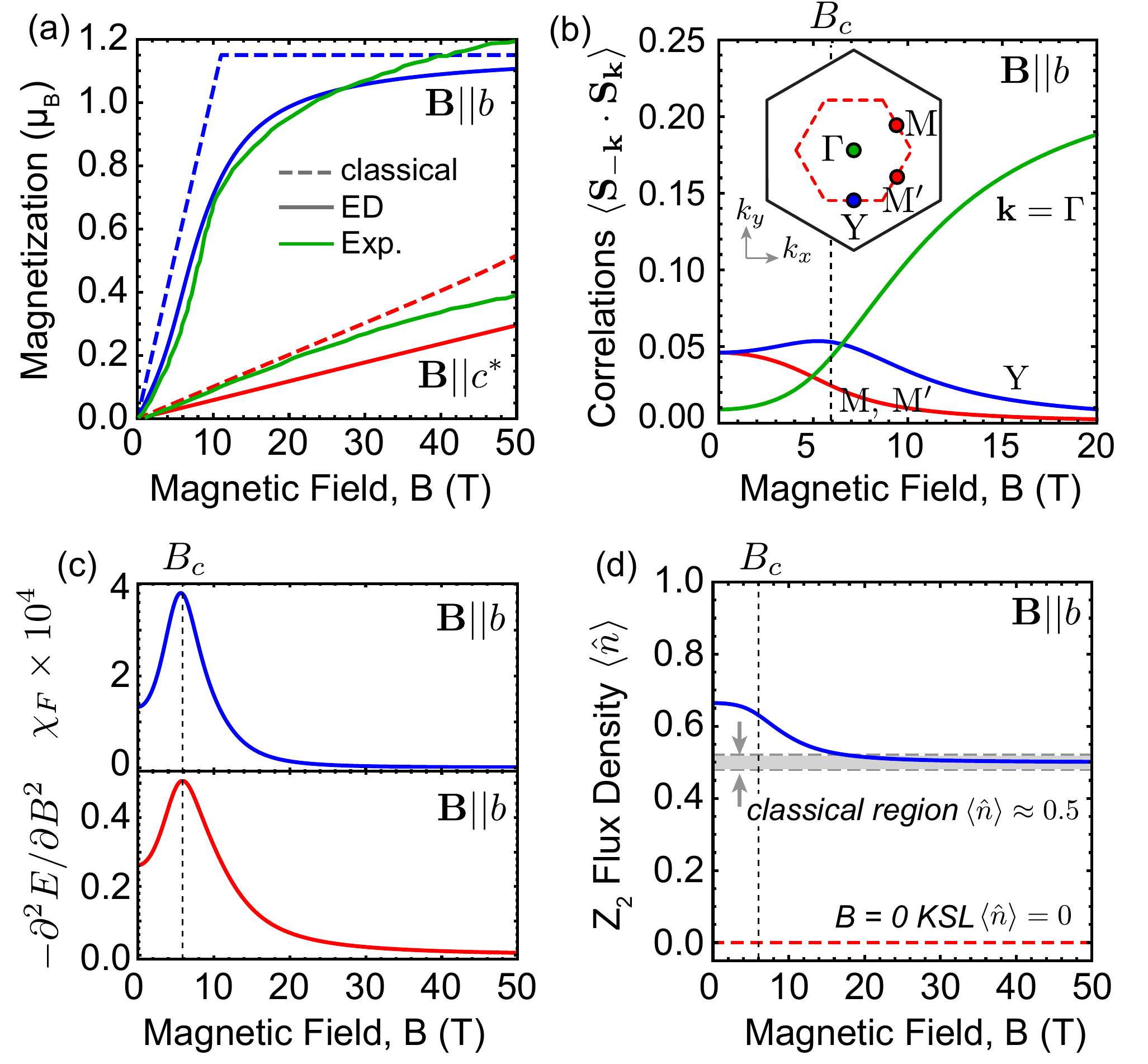}
\caption{Evolution of the $T = 0$ static correlations under magnetic field computed via ED. (a) Magnetization $M(B)$. Experimental data at $T$ = 2 K from \cite{PhysRevB.92.235119}. (b) Static structure factor for $\mathbf{k} = \Gamma$, M, and Y. (c) Ground state fidelity susceptibility $\chi_F$ and second derivative of the ground state energy. The peak in both indicates a single phase transition at $B_c \sim$ 6 T. (d) $\mathbb{Z}_2$ flux density compared to known limits: the Kitaev spin liquid (KSL) has $\langle \hat n \rangle = 0$ at $B = 0$, while classical collinear ordered states have $\langle \hat n \rangle \approx 0.5$. The present model has $\langle \hat n \rangle \gtrsim 0.5$ at all fields (blue line).}
\label{fig-1}
\end{figure}

{\it Results $-$} 
We first discuss the static correlations at zero temperature,
computed via exact diagonalization (ED) on the 24-site
cluster in Fig.~\ref{fig-lat}b for $\mathbf{B}||b$. Results for $\mathbf{B}||a$ are similar and are shown in the Supplemental Material \cite{sup}. The anisotropy in the computed magnetization (Fig.~\ref{fig-1}a) agrees well with experimental data at $T$ = 2 K, thus
providing a consistency check for the present model. At low fields, the static structure factor
$\langle \mathbf{S}_{-\mathbf{k}}\cdot \mathbf{S}_\mathbf{k} \rangle$ is peaked
at the M-, M$^\prime$-, and Y-points, corresponding to the three possible
domains of zigzag order (Fig.~\ref{fig-1}b). 
Application of small fields differentiates the zigzag domains, stabilizing $\mathbf{Q}$ = Y for $\mathbf{B}||b$ and $\mathbf{Q}$ = M,M$^\prime$ for $\mathbf{B}||a$.
For fields $B>B_c \sim 6$ T, the suppression of $\langle \mathbf{S}_{-\mathbf{k}}\cdot \mathbf{S}_\mathbf{k}\rangle$ at the zigzag wavevectors, and growth of correlations at $\mathbf{k}=0$ for both $\mathbf{B}||a,b$, indicates a transition towards a paramagnetic state with a substantial ferromagnetic polarization.

 In principle, this transition may occur directly, or proceed via one or more intermediate states \cite{yadav2016kitaev,janssen2017magnetization,VMC}. For the present model, we resolve only one phase transition at $B_c \sim 6$ T for both $\mathbf{B}||a,b$, as evidenced by a single peak in the second derivative of the ground state energy ($-\partial^2E_0/\partial B^2$) and ground state fidelity susceptibility $\chi_F = \frac{2}{(\delta B)^2}(1-\langle \Psi_0 (B) | \Psi_0 (B+\delta B)\rangle$), shown in Fig.~\ref{fig-1}c. The appearance
 of only one transition indicates that the high-field state is adiabatically connected to the fully polarized state
and is therefore topologically trivial. The finite value of $\chi_F$ at all
fields is consistent with a continuous transition,
suggesting that $T_N$ may terminate in a quantum critical point at $B_c$
\cite{wolter2017field} for both $\mathbf{B}||a,b$. This is in contrast to the
results of a mean-field analysis, which found the transition with
$\mathbf{B}||b$ to be continuous, while for $\mathbf{B}||a$ to be first order
\cite{janssen2017magnetization}. The magnitude of the critical field $B_c\sim
6$ T in ED calculations agrees well with the range of 6 - 8 T observed
experimentally
\cite{PhysRevLett.118.187203,PhysRevB.95.180411,wolter2017field,hentrich2017large,zheng2017gapless,baek2017observation}. The reduction with respect to the classical transition fields of 11 T ($\mathbf{B}||b$) and 8.2 T ($\mathbf{B}||a$) is likely the effect of quantum fluctuations. Similarly, the computed magnetization in ED lies below the classical
value (Fig.~\ref{fig-1}a) at all finite fields. In contrast with pure SU(2) Heisenberg interactions,
the fully polarized state would not be an eigenstate of $\mathcal{H}$, so that quantum
fluctuations reduce the magnetization ($M(B)$) even at high field
\cite{janssen2017magnetization, PhysRevB.92.235119}.

 \begin{figure*}
\includegraphics[width=\linewidth]{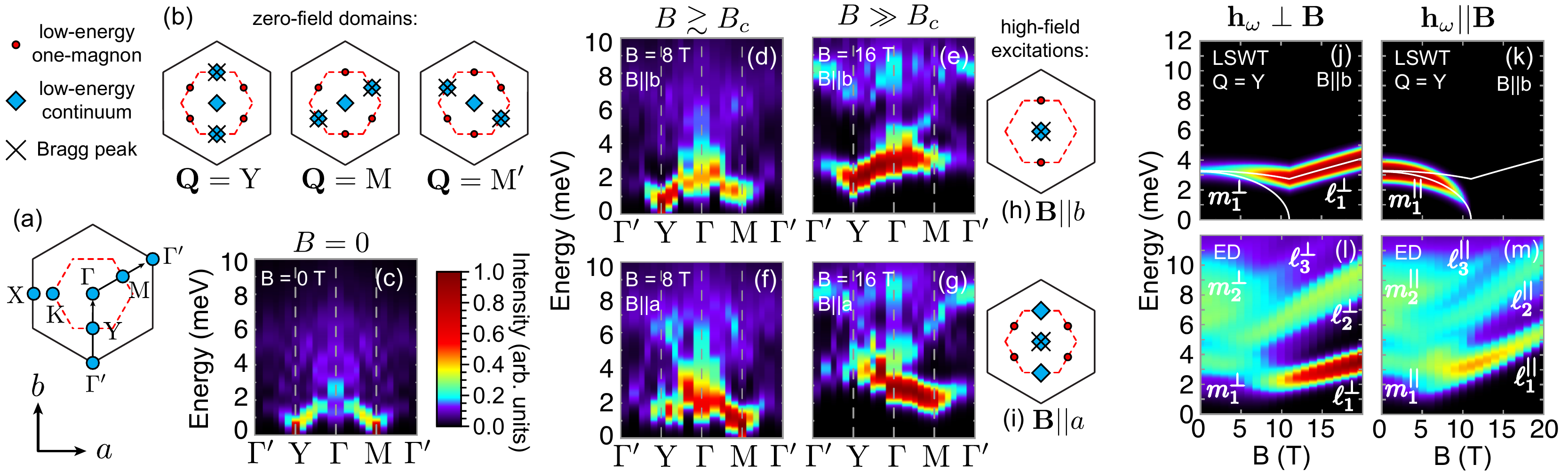}
\caption{(a) Brillouin zone definition. (b) Summary of low-energy contributions to $\mathcal{S}^{\mu\nu}$ from different zigzag domains at $B=0$. (c-g) $T = 0$ inelastic neutron scattering intensity $\mathcal{I}(\mathbf{k},\omega) \propto f(k)^2\sum_{\mu\nu} (\delta_{\mu,\nu} - \hat k_\mu \hat k_\nu)\mathcal{S}^{\mu\nu}(\mathbf{k},\omega)$ under applied field, computed with ED; $f(k)$ is the magnetic form factor for Ru$^{3+}$ \cite{Cromer:a04464}. (c) $B=0$, (d,e) $\mathbf{B}||b$, (f,g) $\mathbf{B}||a$. (h,i) Summary of low-energy contributions to $\mathcal{S}^{\mu\nu}$ for $B>B_c$. (j-m) Polarized electron spin resonance absorption $\omega \chi^{\prime\prime}(\omega) \propto\omega \mathcal{S}^{\mu\mu}(0,\omega)$, with $\mu||\mathbf{h}_\omega$, at the level of (j,k) LSWT and (l,m) ED. 
LSWT results include only the domain $\mathbf{Q}$ = Y for $B<B_c$. ED results combine data from various 20- and 24-site clusters as in \cite{winter2017breakdown}. Spectra were Gaussian broadened by a width of 0.5 meV, and integrated over $k_{c^*}$ consistent with \cite{banerjee2016neutron}. The color scale of each figure is independent. }
\label{fig-sw}
\end{figure*}

In order to further characterize the high- and low-field states, we show, in
Fig.~\ref{fig-1}d, the evolution of the $\mathbb{Z}_2$ flux density appropriate
for the KSL. This is $\langle\hat n\rangle =
\frac{1}{2}(1-\langle\hat{W}_p\rangle)$, where $\hat{W}_p = 2^6
S_1^xS_2^yS_3^zS_4^xS_5^yS_6^z$ (refer to Fig.~\ref{fig-lat}b for site labels).
In the limit of pure $K_1$ interactions and $B=T=0$, the KSL has $\langle\hat
W_p\rangle=+1$ and $\langle\hat n\rangle=0$, signifying the absence of fluxes
\cite{kitaev2006anyons}. In contrast, any classical collinear ordered state must have
$\langle\hat n\rangle \approx \frac{1}{2}$, which would imply both a large
flux density, and a maximum in the variance of the flux density, $\Delta n
= \sqrt{\langle \hat{n}^2\rangle - \langle\hat n\rangle^2} \approx
\frac{1}{2}$. That is, any state
with a sizeable ordered moment cannot have a well-defined $\hat n$, since $[\hat{\mathbf{S}}_i,\hat W_p] \neq 0$.
Numerically, we find that $\langle\hat n\rangle$ indeed reaches $\sim \frac{1}{2}$ at
high field. Interestingly, at low-field, the computed flux density is even {\it
larger} than this classical value. For $\Gamma_1 > 0$, the energy is
minimized for off-diagonal correlations $\langle S_i^\alpha S_j^\beta \rangle <
0$, which effectively enhance $\langle \hat n \rangle$. 

Given the large
$\langle \hat n \rangle$ and $\Delta n$ in the ground state of the present
model at all fields, discussion of the excitations in terms of the fluxes and
spinons of the $\mathbb{Z}_2$ KSL may not provide the most appropriate starting
point at $T = 0$. Consistently, \cite{VMC} found all $\mathbb{Z}_2$ states to
have poor variational energies for a similar model. We therefore choose
the description in terms of  magnon
and multi-magnon (continuum) excitations, which can be understood perturbatively starting from a mean-field description of the zigzag or field polarized state.

An important consequence of the bond-dependent interactions in real space, is
that low-energy contributions to the dynamical structure factor $\mathcal{S}^{\mu\nu}(\mathbf{k},\omega) =
\int dt \ e^{-i\omega t} \langle S^\mu_{-\mathbf{k}}(t)
S^\nu_{\mathbf{k}}(0)\rangle$ appear at locations in $k$-space related to the
polarization $\mu,\nu \in \{x,y,z\}$ \cite{PhysRevB.92.024413}. This
observation applies equally to the present model, and to other ``Klein-dual''
phases \cite{PhysRevB.94.201110,PhysRevB.89.014414,hermanns2017physics}. As a
result, rotation of the local moments $\mathbf{m}_i(B)$ with respect to the
anisotropy axes dramatically restructures the low-energy excitations at finite
$\mathbf{B}$, which can be anticipated at the level of LSWT. Here we use
the LSWT reference (see Fig.~\ref{fig-sw}b,h,i and the Supplemental Material \cite{sup}) to analyze 
the INS intensity $\mathcal{I}(\mathbf{k},\omega)$ computed via ED calculations.

At zero field the ED response (Fig.~\ref{fig-sw}c) reflects a mixture of the three zigzag
domains. We note however that within each domain, the low-energy magnons appear at wavevectors away from the Bragg
peak position and a {\it continuum response is expected near the $\Gamma$-point},
due to a strong and kinematically allowed decay process for the single
magnons \cite{winter2017breakdown}. For example, at $B = 0$, the zigzag domain with Bragg peak at Y has low-energy
magnons at M and M$^\prime$, while low-energy (multi-magnon) continuum states
appear near the Y- and $\Gamma$-points (Fig.~3b). For the latter $k$-points, the extension of the
multi-particle continuum below the single magnon excitations implies the spontaneous decay of magnons, provided
coupling to the continuum is symmetry-allowed
\cite{zhitomirsky2013colloquium,maksimov2016field}, which is the case for the Hamiltonian in (1). For
$\mathbf{B}||b$ and $B>B_c$, the rotation of moments causes the magnons at M and M$^\prime$
to shift to higher energy, while new soft magnons appear at the
Y-point (Fig.~\ref{fig-sw}d,e,h), which is the Bragg peak position of the most stable zigzag domain below $B_c$. Low-energy continuum excitations remain
near the $\Gamma$-point, implying the continuum may remain stable at high field. Analogous effects occur for
$\mathbf{B}||a$  (Fig.~\ref{fig-sw}f,g,i and Supplemental Material~\cite{sup}).
Specifically for $\mathbf{B}||a$ and $B>B_c$, the
lowest-energy magnons appear at M and M$^\prime$, while the
{\it lowest-energy continuum states} appear at Y and $\Gamma$. Together, these
results may explain the observed absence of sharp low-energy magnons at high
field $\mathbf{B}||a$, along the $\mathbf{k}$-path $\Gamma$-Y-$\Gamma^\prime$ (recently
reported in \cite{arnabnew}).

 The composition of this continuum near $k=0$ has been a matter of significant discussion,
as the breakdown of magnons may signify the emergence of unconventional
excitations. To investigate the dynamical response at $k=0$, we show, in Fig.~\ref{fig-sw}(j-m), the ESR response
$\omega\chi^{\prime\prime}(\omega)$ at the level of ED and LSWT for $\mathbf{B}||b$ (results for $\mathbf{B}||a$
are similar \cite{sup}). 
For $B<B_c$, the ESR response should be dominated by the zigzag domain with Bragg
point $\mathbf{Q}$ = Y. At the LSWT level, two intense one-magnon bands are
anticipated, labelled $m_{1}^{||}$ and $m_{1}^{\perp}$ (Fig.~3j,k), with
dominant intensity for oscillating magnetic field $\mathbf{h}_\omega$ polarized
$|| \mathbf{B}$ and $\perp \mathbf{B}$, respectively. These modes also appear
in ED (Fig.~3l,m), with the addition of {\it broad continuum excitations} centered around 6
- 8 meV, labelled $m_{2}^{||}$ and $m_{2}^{\perp}$. The polarization dependence
of $\omega\chi^{\prime\prime}(\omega)$ for $B<B_c$ is likely underestimated in
ED due to the persistence of $\mathbf{Q}$ = M, M$^\prime$ zigzag correlations
resulting from finite-size effects (see Fig.~2b). For fields $B>B_c$, LSWT
predicts only one intense one-magnon $\ell_{1}^{\perp}$ excitation of
transverse ($\mathbf{h}_\omega \perp \mathbf{B}$) polarization, while the ED
response shows multiple excitation branches. The lowest energy mode
$\ell_{1}^{\perp}$ in ED appears only for $\mathbf{h}_\omega \perp \mathbf{B}$
(Fig.~\ref{fig-sw}l). For this mode, the gap increases linearly with applied
field with a rate of $g_{ab}\mu_B\Delta S \approx$ 0.13 meV/T, with $\Delta S =
1$, consistent with a one-magnon excitation as predicted by LSWT. A second intense band
$\ell_{2}^{\perp}$ appears at higher energy with larger slope $\Delta S \approx
2$, consistent with a two-magnon excitation. For longitudinal ($\mathbf{h}_\omega
|| \mathbf{B}$) polarization, the main excitation branches $\ell_{1}^{||}$ and
$\ell_{3}^{||}$ also evolve with $\Delta S \gtrsim 2$, suggesting a similar
multi-magnon origin. Finally, weak higher energy modes $\ell_{3}^{\perp,||}$
also appear with $\Delta S \geq 2$. These results are in qualitative agreement
with recent high-field THz ESR experiments \cite{wang2017magnetic}, offering a
potential interpretation of the observed excitations (for a detailed comparison
see the Supplemental Material~\cite{sup}). In this context, the
application of magnetic field is valuable for `dissecting' the
$\mathbf{k}=\Gamma$ continuum. Such an experimental strategy has recently been
demonstrated for the pyrochlore Yb$_2$Ti$_2$O$_7$
\cite{thompson2017quasiparticle,pan2014low}, which also features anisotropic
bond-dependent interactions.

\begin{figure}
\includegraphics[width=\linewidth]{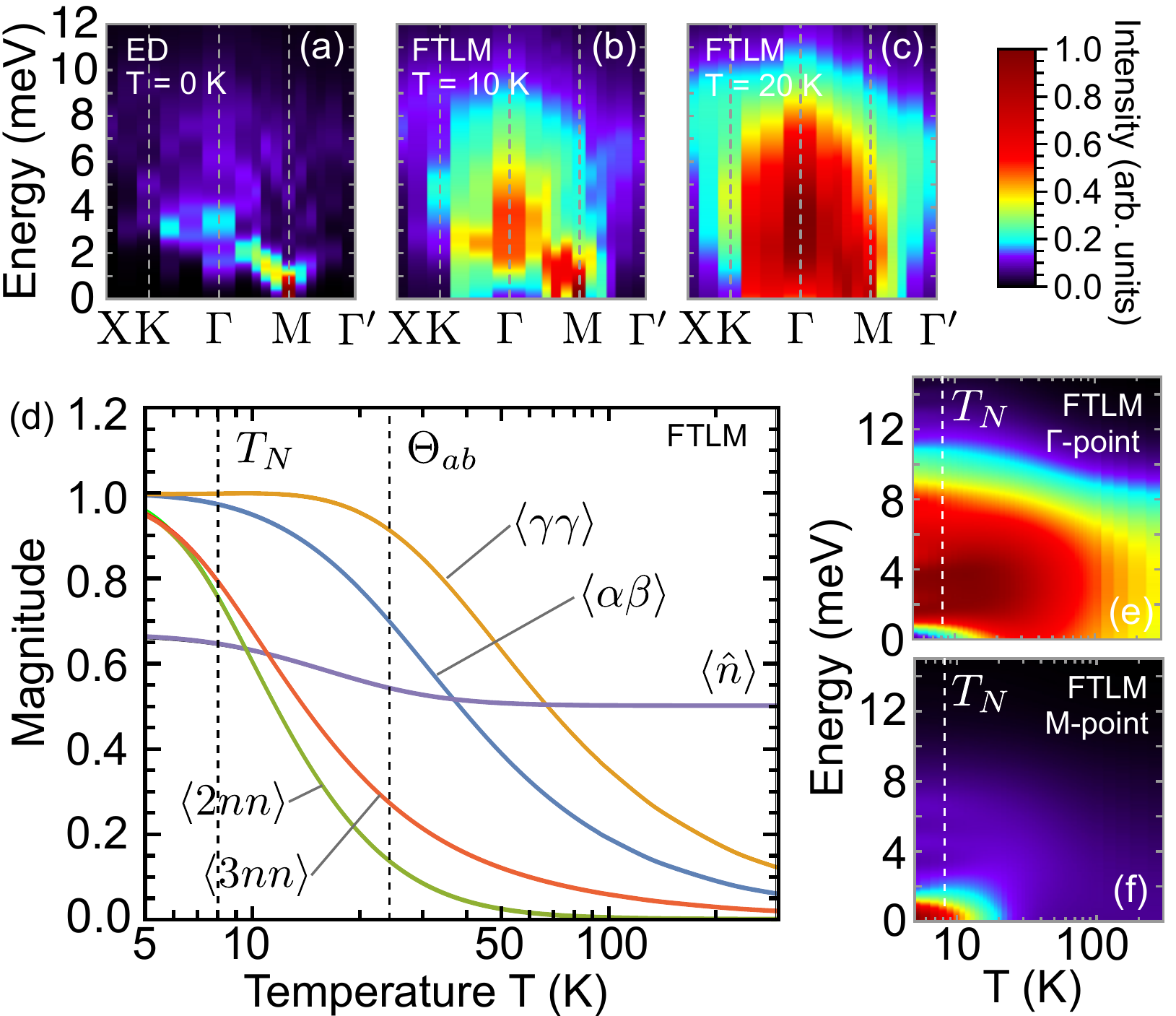}
\caption{Neutron scattering intensity for $T>0$, as a function of $\mathbf{k}$ (a-c) and $T$ for (e) $\mathbf{k}$ = $\Gamma$ and
 (f) $\mathbf{k}$ = M, combining results of multiple clusters. (d): Kitaev flux density $\langle \hat n \rangle$ and normalized real space static correlations computed for the cluster in Fig.~\ref{fig-lat}b, for first nearest neighbours $\langle \gamma\gamma\rangle \equiv \langle S_1^\gamma S_2^\gamma\rangle_T$, $\langle \alpha\beta\rangle \equiv \langle S_1^\alpha S_2^\beta\rangle_T$, second neighbours $\langle 2nn \rangle \equiv \langle \mathbf{S}_1 \cdot \mathbf{S}_3\rangle_T$, and third neighbours $\langle 3nn \rangle \equiv \langle \mathbf{S}_1 \cdot \mathbf{S}_4\rangle_T$. Except for $\langle \hat n\rangle$, values are normalized by their $T=0$ value. Site labels refer to Fig.~\ref{fig-lat}b. The color scale of each figure is independent.}
\label{fig-2}
\end{figure}

Having described the effect of magnetic field on the excitations, we now discuss the effects of finite temperature for $B=0$.
 Results computed via the Finite Temperature Lanczos method (FTLM) \cite{FTLM} are shown in Fig.~\ref{fig-2}. Analysis of statistical errors suggests reliable results for $T \gtrsim 5$ K, see \cite{sup}.
 We first estimate $T_N \approx 8$ K from a maximum in $-(\partial/\partial T) \langle \mathbf{S}_{-\mathbf{k}}\cdot \mathbf{S}_\mathbf{k}\rangle_T$, with $\mathbf{k}$ = M, $\Gamma$. This value is comparable to the experimental values of 7 - 14 K \cite{PhysRevB.95.180411,wolter2017field,banerjee2016proximate}. 
 Upon increasing $T$ above $T_N$ we find a marked shift of the low-energy INS
spectral weight away from the zigzag wavevectors, towards the $\Gamma$-point
(Fig.~\ref{fig-2}e,f), consistent with INS
experiments~\cite{banerjee2016proximate,arnabnew}. Above $T_N$, the $g_{ab} >
g_{c^*}$ emphasizes short-ranged correlations between spin-components in the
$ab$-plane, which are ferromagnetic due to $K_1< 0$ and $\Gamma_1 > 0$. This is
revealed by the positive in-plane Curie-Weiss constant, $\Theta_{ab} \sim
-(3J_1+K_1-\Gamma_1+3J_3)/(4k_B)$, which is $\Theta_{ab} \sim + 22$ K for the
present model (experimentally, $\Theta_{ab} \sim $ +38 to +68 K
\cite{PhysRevB.91.144420,banerjee2016neutron,PhysRevB.91.180401}). For this
reason, the suppression of zigzag order for $T>T_N$ is expected to generate
dominant scattering intensity at $k=0$, reflecting the emergence of
short-ranged ferromagnetic correlations. Overall, the finite temperature
spectra agree well with experimental INS observations\cite{do2017incarnation},
suggesting that the present model may also capture the essential features of
the dynamics above $T_N$.

An interesting question therefore remains to what extent this region $T_N < T < \Theta$  (Fig.~\ref{fig-lat}a) 
can be connected to the response of the pure Kitaev model, given the evidence for
large $\Gamma_1$ interactions in $\alpha$-RuCl$_3$.  For purely Kitaev
interactions, the intermediate $T$ regime would be characterized by a large
density of thermally excited
fluxes \cite{nasu2015thermal,yoshitake2017majorana}, which likely confine the
fermionic spinons \cite{samarakoon2017comprehensive,hermanns2017physics}. This
regime is characterized by a saturation of nearest neighbour spin-spin
correlations. For the present model, we find deviations from Curie-Weiss
behaviour below $T\sim 70$ K, while nearest neighbour correlations saturate for
$T \lesssim \Theta_{ab}$ (Fig.~\ref{fig-2}d). Longer range correlations set in
near $T_N \sim 8$ K, suggesting the intermediate temperature regime may be
relatively narrow. If the ordering of fluxes at low temperatures is preempted
by magnetic order, then a deconfined region may not appear. Consistent with
this picture, we find that the Kitaev flux density remains $\langle \hat n
\rangle \gtrsim \frac{1}{2}$ at all temperatures for the present model
(Fig.~\ref{fig-2}d). This leaves two possibilities for the intermediate
temperature dynamics. Either, {\it all} correlations are short-ranged,
suggesting the phase cannot be qualitatively distinguished from a conventional
paramagnet, or there exist higher order long-range or algebraic spin
correlations. These could be associated with alternative quantum ground states
suggested for finite $\Gamma_1$ interactions
\cite{catuneanu2017realizing,ybknew,PhysRevLett.118.147204}, which are not
characterized by $\langle \hat n \rangle$. In this sense, development of probes
for higher order correlations (such as RIXS \cite{halasz2016resonant}) may
prove vital for further understanding the intermediate $T$ regime.
Investigating the $T>0$ classical dynamics \cite{samarakoon2017comprehensive}
of the full $(J_1,K_1,\Gamma_1,J_3)$ model also represents an important avenue
of future study. 

{\it Conclusions $-$} We have shown that the model for $\alpha$-RuCl$_3$ defined in Eq.~(1) reproduces many key aspects of the experimental observations, including the relevant energy scales ($B_c$ and $T_N$), and the evolution of the dynamical response at finite $T$ and $\mathbf{B}$. In the range of $T$ and $B$ studied, we do not find any regime where the $\mathbb{Z}_2$ fluxes of the Kitaev form ($\hat W_p$) are dilute, which hampers possible connections to Kitaev's exact solution. 
We find the high-field phase to be smoothly connected to the fully polarized state.
Nonetheless, the evolution of high field excitations reveals significant multiparticle character in the $\Gamma$-point continuum, providing insight into recent ESR experiments. Combined, these results supply a valuable framework for interpreting a wide range of recent studies of $\alpha$-RuCl$_3$.

\begin{acknowledgments}
{\it Acknowledgements $-$ }  We acknowledge useful discussions with A. Banerjee, C. Batista, A. L. Chernyshev, Y. B. Kim, D. Kovrizhin, J. Knolle, A. Loidl, R. Moessner, and S. E. Nagler. S. M. W. acknowledges support through an NSERC Canada
Postdoctoral Fellowship. R. V. and K. R. acknowledge support by the
Deutsche Forschungsgemeinschaft through grant SFB/TR 49 and the computer time was allotted by the Centre for for Scientific Computing (CSC) in Frankfurt. R.C. was supported in part by EPSRC under Grant No. EP/M020517/1
and by KITP under NSF Grant No. PHY11-25915.
\end{acknowledgments}

\bibliography{magfield}

\clearpage



\section*{Supplementary material}

# Supplemental Information for: Probing $\alpha$-RuCl$_3$ Beyond Magnetic Order: Effects of Temperature and Magnetic Field

Stephen M. Winter*,[1] Kira Riedl,[1] David Kaib,[1] Radu Coldea,[2] and Roser Valentí[1]

[1] *Institut für Theoretische Physik, Goethe-Universität Frankfurt, Max-von-Laue-Strasse 1, 60438 Frankfurt am Main, Germany*

[2] *Clarendon Laboratory, University of Oxford, Parks Road, Oxford OX1 3PU, United Kingdom*

(Dated: February 7, 2018)

**Please Note:** As noted in the main text, the $g$-values employed throughout ($g_{ab} = 2.3$ and $g_{c*} = 1.3$) are consistent with the range of previous theoretical estimates for $\alpha$-RuCl$_3$ [1, 2], and experimental values for similar compounds [3–5].

The initially submitted version of this manuscript contained an additional supplemental section further justifying such choice, based on estimates of the $g$-tensor associated with a total trigonal crystal field splitting estimated to be $\sim 50$ - 60 meV. However, we have subsequently noted an unusually large method- and structure-dependence in ab-initio estimates of this crystal field splitting. This enlarges the uncertainties in the $g$-value estimates when combined with the strong sensitivity of the $g$-values to such computed parameters. In order to avoid confusion, we have therefore removed this section in subsequent versions.

This version of the supplemental material reflects the final published version.

## S.1: Further Exact Diagonalization Results

### S.1.1: Static Correlations for $\mathbf{B}||a$ and $T = 0$

A comparison of the static correlations for $\mathbf{B}||a$ and $\mathbf{B}||b$ is shown in Fig. S1. As noted above, a $g$-value of $g_{ab} = 2.3$ has been employed in both cases. At the mean-field (classical) level, the transition to the field-polarized state is first order for $\mathbf{B}||a$, as indicated by a jump in the magnetization in Fig. S1a, at approximately 8.2 T. In ED calculations, this jump is absent, which may indicate either a significant influence of finite-size effects, or a modification of the transition order by quantum fluctuations. Under applied field $\mathbf{B}||a$, the static structure factor $\langle \mathbf{S}_{-\mathbf{k}} \cdot \mathbf{S}_{\mathbf{k}} \rangle$ is suppressed at the Y-point faster than at the M- and M′-points (Fig. S1b). In general, the zigzag domains with local moments oriented perpendicular to the applied field at $B = 0$ tend to remain stable at finite field. The combination of $K_1$ and $\Gamma_1$ interactions confine the local moments to lie in the plane perpendicular to the zigzag propagation vector. Thus, $\mathbf{Q} = $ Y tends to be uniquely stabilized for $\mathbf{B}||b$, while $\mathbf{Q} = $ M and M′ are preferred for $\mathbf{B}||a$. Overall, the magnetization, $\chi_F$, $\partial^2 E/\partial B^2$, and $\langle \hat{n} \rangle$ computed via ED all behave similarly for $\mathbf{B}||b$ and $\mathbf{B}||a$.

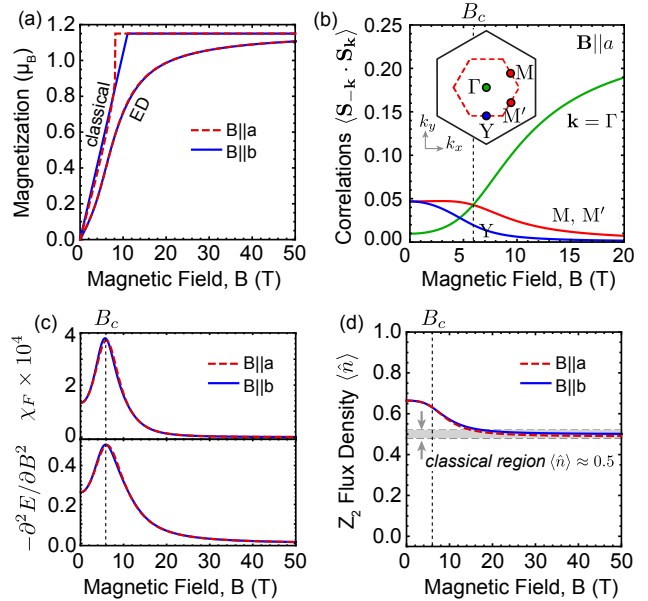

FIG. S1. Comparison of the $T = 0$ static correlations under magnetic field computed via ED for $\mathbf{B}||a, b$. (a) Magnetization $M(B)$. (b) Static structure factor for $\mathbf{B}||a$. Results for $\mathbf{B}||b$ appear in the main text. (c) Ground state fidelity susceptibility $\chi_F$ and second derivative of the ground state energy. (d) $\mathbb{Z}_2$ Kitaev flux density.

### S.1.2: ESR Response for $\mathbf{B}||a$.

The computed ESR response for $\mathbf{B}||a$ is shown in Fig. S2. Results were obtained as described in the main text. At the LSWT level, the response shows an abrupt change at the first order transition. The ED results show a continuous evolution, with multiple excitation branches emerging at high field, similar to the case of $\mathbf{B}||b$, presented in the main text.

For the ED results for $\mathbf{B}||a$, an additional mode with $\Delta S = 1$ appears for $\mathbf{h}_\omega \perp \mathbf{B}$, which is labelled $\ell_*^\perp$ in Fig. 2c. The appearance of this mode may reflect a continuation of the $m_1^\perp$ mode into the high-field state due to the anomalous persistence of zigzag correlations above $B_c$ in the ED calculations (due to finite size effects). The strong suppression of the intensity of this mode with increasing field is suggestive of this assignment.

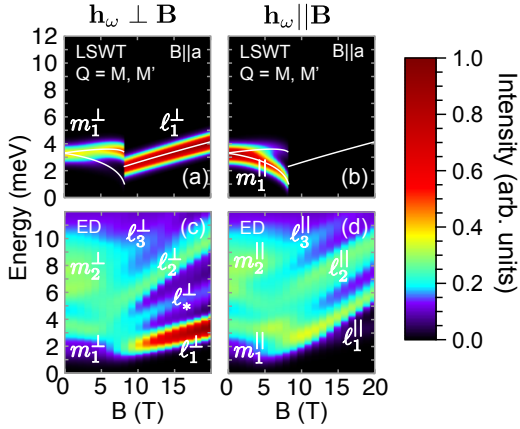

**FIG. S2.** Polarized electron spin resonance absorption $\omega\chi''(\omega) \propto \mathcal{S}^{\mu\mu}(0,\omega)$, with $\mu||\mathbf{h}_\omega$, for $\mathbf{B}||a$. Results are for (a,b) LSWT and (c,d) ED.

## S.2 Detailed comparison of ED results with high-field THz ESR experiments

In this section we compare our numerical results with recent high-field THz ESR experiments by Wang et al. [6] and Ponomaryov et al. [7], and propose an identification of the experimentally observed excitations. The former measurements are polarization resolved, while the latter measurements were performed in the Voigt geometry, and therefore should contain both excitations with $\mathbf{h}_\omega||\mathbf{B}$ and $\mathbf{h}_\omega \perp \mathbf{B}$.

- One-magnon modes $(B < B_c)$: In the zigzag ordered phase of the presently studied model, there are two different one-magnon excitations that appear below 4 meV within LSWT (Fig. 3j,k and Fig. S2a,b). The lowest energy magnon is excited for $\mathbf{h}_\omega||\mathbf{B}$, and has decreasing intensity on approaching $B_c$. The higher energy magnon is excited for $\mathbf{h}_\omega \perp \mathbf{B}$, and has increasing intensity on approaching $B_c$. These excitations are labelled $m_1^\perp$ and $m_1^{||}$ in Fig. 3 and S2. The excitation energy of the $m_1^{||}$ mode decreases sharply with increasing field, while the energy of the $m_1^\perp$ mode is less field-dependent.

  Comparing to the experimental data, these two modes appear consistent with $M_{1,||}$ and $M_{1,\perp}$ from Ref. [6] and modes D and E from Ref. [7]. Thus we identify such excitations as one-magnon modes, which are "dressed" by symmetry-allowed coupling to the multi-magnon continuum, which extends below the one-magnon energies both at zero field [8], and near $B_c$ (Fig. S3).

- Continuum contributions $(B < B_c)$: At higher energy additional continuum contributions appear in the ED results (Fig. 3l,m and Fig. S2c,d), which

are not associated with any one-magnon excitations of LSWT. These appear around 6 meV in the present ED calculations and are labelled $m_2^\perp$ and $m_2^{||}$ in Fig. 3 and S2.

These excitations appear consistent with the behaviour of features $M_{2,||}$ and $M_{2,\perp}$ from Ref. [6]. Such continuum contributions are also resolved in the neutron scattering experiments [9, 10]. However, these broad features may not have been clearly resolved in the experiments of Ref. [7].

- One-magnon mode $(B > B_c)$: At the level of LSWT, a single intense one-magnon excitation is expected to appear for $B > B_c$ exclusively in the polarization $\mathbf{h}_\omega \perp \mathbf{B}$. A similar excitation appears also in ED, and is labelled $\ell_1^\perp$ in Fig. 3 and S2. As observed in the ED results, this mode would be expected to become increasingly sharp and intensive with increasing field, due to suppression of the coupling to the multi-magnon continuum.

  This mode is consistent with the $L_{1,\perp}$ mode of Ref. [6], and likely corresponds to the A mode of Ref. [7]. We do not resolve any additional weak F mode observed in Ref. [7].

- Two-magnon excitations $(B > B_c)$: For $\mathbf{B}||b$ and $\mathbf{B}||a$, there appear additional modes with polarization $\mathbf{h}_\omega \perp \mathbf{B}$, which have a larger slope than the putative one-magnon excitation. One of these modes persists with significant ESR intensity to the highest fields probed. As we argue in the main text, this is likely a two-magnon excitation.

  We assign this to the $L_{2,\perp}$ mode of Ref. [6], and have labelled it $\ell_2^\perp$ in Fig. 3 and S2. This may correspond to the B and/or C modes of Ref. [7].

  For $B > B_c$, there are two main excitations in the polarization $\mathbf{h}_\omega||\mathbf{B}$. The lowest one shows increasing intensity over the field range 10 - 15 T, where it splits away from the $\ell_1^\perp$ mode. We assign this excitation to the $L_{1,||}$ excitation of Ref. [6], and label it $\ell_1^{||}$ in Fig. 2 and S2. It is not clear whether this excitation is resolved in Ref. [7], but may appear under the A mode for much of the measured field-range.

  The higher energy excitation (labelled $\ell_2^{||}$) appears to correspond to the $L_{2,||}$ mode of Ref. [6], and the B and/or C modes of Ref. [7]. It should be noted that the $\ell_2^{||}$ and $\ell_2^\perp$ excitations are distinct, even though they have similar excitation energies.

We finally note that these proposed assignments are motivated primarily by the relative energies and polarization of the experimentally observed excitations. As shown in Fig. S3, the field dependence of the excitation

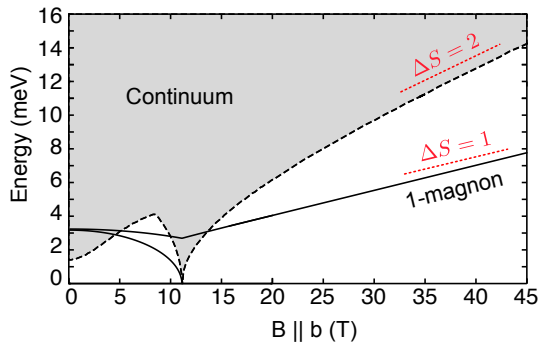

FIG. S3. Schematic evolution of the lowest-energy one magnon (solid lines) and continuum excitations (shaded region with dashed lower boundary) at **k** = 0 based on LSWT for **B**||$b$. At low field (and near $B_c$), the mixing of the continuum and one-magnon states may significantly alter the curvature of the excitation energies. At high field, the slopes are expected to asymptotically approach $\Delta E/\Delta B = g\mu_b\Delta S$, with $\Delta S = 1$ for one-magnon modes, and $\Delta S = 2$ for the two-magnon excitations forming the bottom of the multi-magnon continuum.

energies predicted by LSWT (for **B**||$b$) may show significant curvature near $B_c$, where the closure of the one-magnon gap at the Y-points implies a closure of the two-magnon gap at **k** = 0. Near $B_c$, these slopes may be further altered due to the interaction of the one-magnon and continuum excitations which is not considered in LSWT. As a result, the slope $\Delta E/\Delta B$ is only expected to approach $g_{ab}\mu_b\Delta S$ in the asymptotic limit of large field. Very near to $B_c$, the curvature of the excitation energies is unlikely to be correctly captured in ED as finite size effects prohibit the closure of the excitation gaps. A further experimental test of the proposed assignments would therefore be to measure the evolution of each excitation at large field $B \gg B_c$, on approaching the asymptotic limit.

## S.3: Location of Low-Energy Excitations at Finite Magnetic Field

### S.3.1: Evolution of the Excitations for **B**||$b$ and **Q** = Y at the LSWT Level

For **B**||$b$, the zigzag domain with **Q** = Y is preferred, as the zero-field ordered moments $\mathbf{m}_i(0)$ lie in the $ac^*$-plane, such that $\mathbf{m}_i(0) \perp \mathbf{B}$. It is therefore relevant to consider the evolution of the neutron scattering intensity at the level of linear spin wave theory (LSWT) for the domain with **Q** = Y, shown in Fig. S4. This approach allows for qualitative statements about the location of intense one-magnon contributions to the spectra, as well as low-energy two-magnon continuum states.

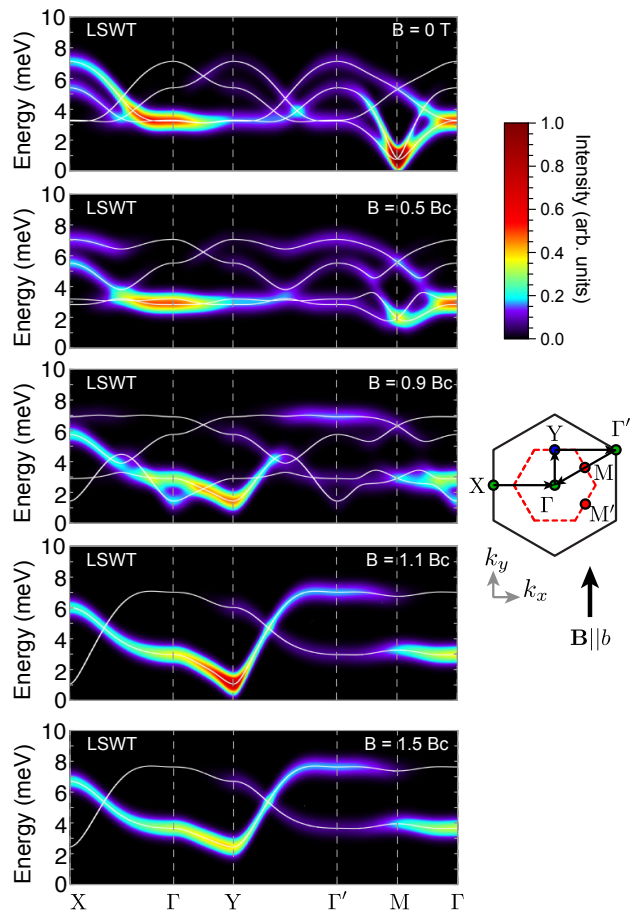

FIG. S4. Evolution of the neutron scattering intensity predicted within LSWT for a single domain with **Q** = Y, and **B**||$b$. Solid lines indicate the one-magnon dispersion branches. At low field, soft modes appear at the M and M′ points. At high field, soft modes appear only at the Y-point. Below $B_c$, twice as many dispersion bands occur due to the doubling of the magnetic unit cell by the spontaneous zigzag AFM order. $B_c = 11$ T.

At low field, soft one-magnon excitations appear at M and M′, corresponding to the pseudo-Goldstone modes of the zigzag order. Upon increasing field **B**||$b$, these shift rapidly to higher energy, while new soft modes appear near Γ and Y. It should be noted that, within the zigzag ordered phase with **Q** = Y, the Γ- and Y-points are related by a reciprocal lattice vector of the magnetic unit cell. Thus, modes appearing at Γ and Y must be energetically degenerate within the zigzag ordered phase, but typically carry different spectral intensities. Upon approaching $B_c$ with increasing field, the spectral intensity of the lowest Γ-point excitation vanishes. This mode is only present in the zigzag phase where it appears due to the doubling of the magnetic unit cell compared to the high-field paramagnetic state. In the high-field phase, soft one-magnon modes therefore appear only at Y, with polarization parallel to the zero-field ordered moments.

This last result can be anticipated as follows. Within the mean-field picture, the spontaneous development of zigzag order upon lowering the field to $B_c$ occurs via gap closing at the Y point, where zigzag Bragg peaks develop in the ordered phase. Since the soft modes at the Y-point are responsible for restoring the zigzag order, their polarization must coincide with the spontaneously zigzag-ordered moments appearing below $B_c$. Upon including quantum fluctuations beyond mean field, one would expect these qualitative features to remain unchanged, i.e. the transition to occur via softening of the gap at the Y-point of a one-magnon mode with the same polarization as that of the magnetic Bragg peaks that set in immediately below $B_c$. The transition would be expected to occur at a lower field value than predicted classically as quantum fluctuations generically tend to suppress the long-range order.

Due to quantum fluctuations being present at all fields, significant multi-particle scattering continua may appear. At each field, the lowest energy multi-particle continuum excitations take the form of two-magnon excitations, at the LSWT level. At zero-field, a pair of pseudo-Goldstone magnons can have total momentum M − M = M′ − M′ = Γ, or M - M′ = Y. Thus low-energy two-magnon excitations appear at Y and Γ. At high-field $B > B_c$, low-energy magnons appear only at Y. Therefore, low-energy two-magnon excitations appear only at Γ. It can be expected that multi-particle scattering is strongly enhanced near $B_c$ due to the continuous (quantum critical) nature of the field-driven phase transition, with the one-magnon gap expected to close at Y when $B = B_c$ with a gapless continuum at $\mathbf{k} = 0$ made of two- and other even-particle states.

### S.3.2: Evolution of the Excitations for $\mathbf{B}||a$ for $\mathbf{Q} = $ M, M′ at the LSWT Level

For $\mathbf{B}||a$, the zigzag domains with $\mathbf{Q} = $ M and M′ are degenerate, and stabilized with respect to $\mathbf{Q} = $ Y. We therefore show below the evolution of the neutron scattering intensity at the level of linear spin wave theory (LSWT), averaging over the two domains with $\mathbf{Q} = $ M and M′. The combination of multiple stable domains and $\mathbf{B} \nparallel \mathbf{Q}$ make the excitation spectra somewhat more complex than in the previous section. Moreover, the transition for $\mathbf{B}||a$ is first order at the mean-field level, which results in a discontinuous change in the LSWT results at the critical field $B_c \approx 8.2$ T.

At low field, the two stable domains contribute soft modes at all of Y, M, and M′. The domain with $\mathbf{Q} = $ M is associated with pseudo-Goldstone modes at Y and M′, while the domain with $\mathbf{Q} = $ M′ has pseudo-Goldstone modes at Y and M. Experimentally, the application of a small field $\mathbf{B}||a$ of around 2 T suppresses the domain with $\mathbf{Q} = $ Y, resulting in a relative enhancement of the inelas-

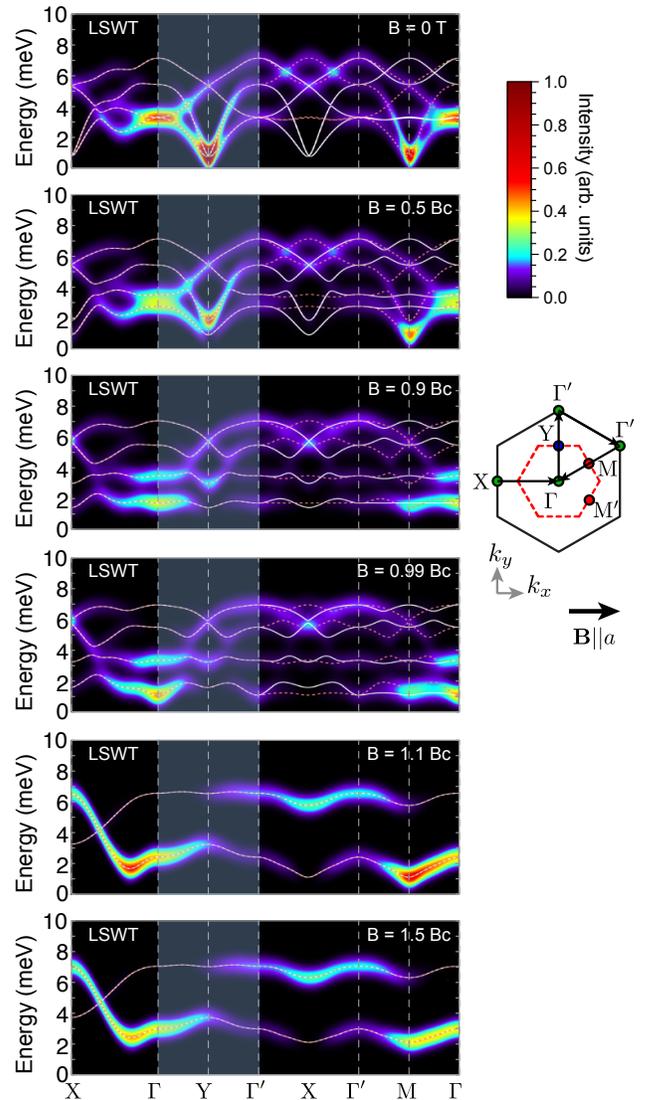

FIG. S5. Evolution of the neutron scattering intensity predicted within LSWT averaged over domains with $\mathbf{Q} = $ M, M′ for $\mathbf{B}||a$. Modes associated with $\mathbf{Q} = $ M and M′ are indicated by solid white and dashed pink lines, respectively. At high field, the reduced number of bands reflects the reduced magnetic unit cell (2 sites per primitive cell) in the field-polarized state. At low field, soft modes appear at all of the Y, M, and M′ points. At high field, soft modes appear in the two-dimensional dispersion surface at M, M′. The shaded region indicates the $\mathbf{k}$-path probed in [9]. $B_c = 8.2$ T.

tic intensity around the Y-point [9]. This enhancement is a natural consequence of the fact that zigzag domains with Bragg peaks at M and M′, both have soft modes at Y, so an enhancement of intensity is expected at the Y points at low fields.

Upon further increasing the field, the soft modes at the Y-point shift to higher energy. Soft modes at the M-point (associated with the $\mathbf{Q} = $ M′ domain) also shift to high energies, while modes at M associated with $\mathbf{Q}$

= M shift to lower energies. Interestingly, for fields just below the first order transition $B \lesssim B_c$, an additional soft one-magnon mode is predicted at $\Gamma$. This may be related to the fact that the magnetic field direction is not perpendicular to the zero-field moment directions ($\mathbf{B} \not\perp \mathbf{m}_i(0)$), which results in a finite magnetic torque [11]. That is, the magnetization within each domain is not parallel to the external field for $B \lesssim B_c$. The high-field state must therefore be reached through both a uniform and staggered rotation of the local moments. As a result of the required uniform rotation, intense soft modes must appear at $\Gamma$, in addition to M, and M'. At the LSWT level, the modes at $\Gamma$ are found to be the most intense for $B \lesssim B_c$. This picture naturally explains the observation of recent INS experiments with $\mathbf{B}||a$ that the spin-wave mode near Y "disappears" upon approaching the critical field [9]. Indeed, fluctuations at the Y-point play no part in the mechanism of the transition, as no magnetic Bragg peak appears at Y for $B \lesssim B_c$. As noted in the main text, the experiment of [9] only probed excitations along the path $\Gamma$ - Y - $\Gamma'$, which would miss the intense soft one-magnon excitations expected at M and M', which are more closely related to the mechanism of the transition. For $B \gg B_c$, the appearance of low-energy magnons at both M and M' implies the lowest energy two-magnon continuum states appear near M − M = $\Gamma$, and M − M' = Y.

We note that since the transition is first-order at the mean-field level, no gap is expected to close at $B_c$ coming from either side. However, ED results have suggested that quantum fluctuations beyond the mean-field model may drive the actual transition continuous. In that case one might expect a gap softening at the M and M' points rather than the discontinuous mean-field behaviour. Due to finite size effects, the evolution of the gap cannot be reliably determined from the present ED calculations. It would therefore be interesting if future INS experiments could also probe the scattering near the M and M' points for $\mathbf{B}||a$ to test the prediction that a relatively intense soft mode exists there above the transition, and explore whether the M-point gap closes and reopens as the field passes through $B_c$.

### S.3.3: Concrete Example of Shifting Soft Modes

The specific shifting of the soft modes in the zigzag state with applied field can be understood by considering a concrete example. We therefore consider the excitations within the zigzag phase that occurs at $J_1 = -1$, $K_1 = +2$, and $\Gamma_1 = J_3 = 0$. Despite possessing different signs of the interactions, and no $\Gamma_1$ interactions, this hidden SU(2) point can be smoothly connected to the model for $\alpha$-RuCl$_3$ discussed in the main text. As a result, it shows a similar evolution of the $k$-position of low-energy modes under applied field. (The connecting path proceeds via finite $J_3$, where the zigzag regions for $K_1 < 0$ and $K_1 > 0$ are linked.)

As discussed in [12], the SU(2)-symmetric point can be mapped on to an antiferromagnetic Heisenberg model by the application of a four-sublattice transformation. Members of the same sublattice are linked by third neighbour bonds, such that each of the four sublattices forms an enlarged honeycomb network. Sublattices 1, 2, and 3 are coupled to sublattice 4 via nearest neighbour X-, Y-, and Z-bonds, respectively. The sublattice transformation is then defined as:

$$\text{sublattice 1: } (x', y', z') = (x, -y, -z) \tag{1}$$
$$\text{sublattice 2: } (x', y', z') = (-x, y, -z) \tag{2}$$
$$\text{sublattice 3: } (x', y', z') = (-x, -y, z) \tag{3}$$
$$\text{sublattice 4: } (x', y', z') = (x, y, z) \tag{4}$$

with $J_1 = -1, K_1 = +2$ (original model) and the transformed interactions $J_1' = +1, K_1' = 0$ (dual model). In reciprocal space, the four-sublattice transformation carries a different momentum shift for each spin component, such that spin operators satisfy $S'_{\mu, \mathbf{k}} = S_{\mu, \mathbf{k} + \mathbf{q}_\mu}$ with $\mu = \{x, y, z\}$. Here, the momentum shifts are $\mathbf{q}_x = $ M, $\mathbf{q}_y = $ M', and $\mathbf{q}_z = $ Y. As explained in [12], this allows one to rationalize the location of low-energy excitations in the zigzag state of the original model.

We first repeat the exposition given in [12] for the zero-field case: Consider a zigzag state in the original model with ordering vector $\mathbf{Q} = $ Y, such that the ordered moments are oriented along the cubic $z$-axis. After transformation, this configuration corresponds to a Néel state in the dual model, with ordered moments also along the cubic $z$-axis. This dual Néel state displays a Bragg peak at $\Gamma'$ and low-energy magnons at $\mathbf{k} = \Gamma$ and $\Gamma'$. The Bragg peak corresponds to static $S_z$ correlations, while the magnons are excited by transverse $S_x$ and $S_y$ operators. Transforming back to the original model, the Bragg peak is shifted to $\mathbf{Q} = $ Y, while the low-energy magnons shift to M and M'. For this reason, the low-energy magnons do not appear at the Bragg peak position in the zigzag state at low field. Such a discrepancy between the location of the magnetic Bragg peaks and the soft magnon wavevectors has also been noted in the Kitaev-Heisenberg $(K_1, J_1)$ model for $K_1 < 0$ and $J_1 > 0$ when the ground state has stripy antiferromagnetic order [13]. Such an effect is expected to occur generically for bond-dependent anisotropic couplings, which separate the correlations associated with each spin-component in both real space and reciprocal space. The polarization of the excitations is therefore directly connected to their location in $k$-space.

These results can be extended to the high-field configuration, with $\mathbf{B}||b$, where $b = [\bar{1}10]$ in cubic coordinates. In the original model, the spins would therefore be completely aligned along the $(-1, 1, 0)$ direction. Applying the four-sublattice transformation produces a pla-

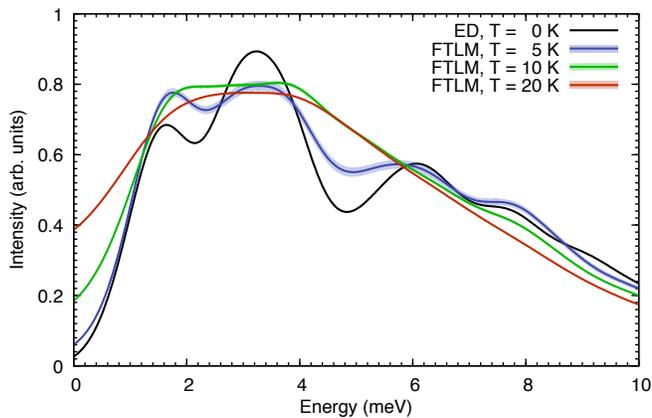

FIG. S6. Neutron scattering intensity at the $\Gamma$-point for various temperatures. For $T > 0$, shaded areas indicate error estimates (if not visible, error regions are thinner than line widths).

nar four-sublattice structure, with moments oriented in the $xy$-plane described by: sublattice 1: $(-1, -1, 0)$, sublattice 2: $(1, 1, 0)$, sublattice 3: $(1, -1, 0)$, and sublattice 4: $(-1, 1, 0)$. This dual configuration must have a low-energy magnon that restores the Néel antiferromagnetic component along the $z$-axis. This magnon lives at $\mathbf{k} = \Gamma'$ in the dual state but is now excited by the $S_z$ operator. Transforming back to the original model, we find that this new low-energy magnon appears at $\mathbf{Q} = \mathrm{Y}$ at high field. Thus, a reorientation of the local moments with respect to the anisotropic couplings results in a shifting of the soft modes.

## S.4: Further Details of ED and FTLM calculations

For $T = 0$, static properties were obtained using the Lanczos algorithm. Dynamical correlation functions by employing the continued fraction method [14] on various 20- and 24-site clusters and combining the data as described in [8].

For $T > 0$, we used the Finite Temperature Lanczos Method (FTLM) [15], with $M = 115$ Lanczos steps. For $\mathcal{I}(\mathbf{k}, \omega)$, we used $R = 200$ random sampling vectors for each of the considered 20- and 24-site clusters. The static correlations, shown in Fig. 4d in the main text, were calculated with $R = 1000$ and employed only the cluster shown in Fig. 1b of the main text. Statistical errors associated with the finite $R$ were estimated by applying

a jackknife resampling technique [16], and are shown in Fig. S6, for the example of $\mathcal{I}(\Gamma, \omega)$. Error regions indicate one standard deviation. For the static correlations presented in Fig. 4d of the main text, the error bars would be thinner than the line widths throughout the shown temperature range.



\end{document}